# Dark Energy From Vacuum Fluctuations

S. G. Djorgovski [a] and V.G. Gurzadyan [b,c]

[a] Division of Physics, Mathematics, and Astronomy, California Institute of Technology
   MS 105-24, Pasadena, CA 91125, USA

[b] ICRANet, ICRA, Physics Dept., University of Rome "La Sapienza", 00185 Rome, Italy

[c] Yerevan Physics Institute, 2 Alikhanyan Brothers St., Yerevan 375036, Armenia

We describe briefly a novel interpretation of the physical nature of dark energy (DE), based on the vacuum fluctuations model by Gurzadyan & Xue, and describe an internally consistent solution for the behavor of DE as a function of redshift. A key choice is the nature of the upper bound used for the computation of energy density contributions by vacuum modes. We show that use of the comoving horizon radius produces a viable model, whereas use of the proper horizon radius is inconsistent with the observations. After introduction of a single phenomenological parameter, the model is consistent with all of the curently available data, and fits them as well as the standard cosmological constant model, while making testable predictions. While some substantial interpretative uncertainties remain, future developments of this model may lead to significant new insights into the physical nature of DE.

## 1. Introduction

The nature of the dark energy (DE) is one of the most outstanding problems of physical sciences today. Numerous and often ingenious attempts have been made to explain it, with multiple new papers appearing daily, yet no model proposed so far has gained a general acceptance.

Modern approaches to the problem date from the pioneering papers by Zel'dovich [1], who was first to recognize the fundamental connection between the quantum physics and the macroscopically observable energy density of the vacuum, $\rho_{vac}$, manifested, e.g., as the cosmological constant. However, the simplest approaches using the Planck energy density lead to the well known problem of being off by some 123 orders of magnitude; and while many suggestions have been made as to how to solve this problem, none are yet compelling.

Gurzadyan & Xue [2,3] (GX) proposed a model in which the energy density is contributed by the quantum fluctuations of the vacuum ground state, rather than the mean energy level itself. Whatever the value of the ground level is (including zero), if it is observed in a finite volume, e.g., within the particle horizon, it will be a subject to quantum fluctuations. Unless there is some as yet unknown limit to applicability of quantum mechanics (which would be a fundamental revelation by itself), such fluctuations are inevitable. Note that *these are vacuum fluctuations, not particle fluctuations*. This approach has been already hinted at by Zel'dovich [1]. Similar considerations have been also discussed recently by Padmanabhan [4,5].

Gurzadyan & Xue [2,3] derived a formula for the energy density contributed by the fluctuations, making reasonable assumptions of a simple topology, homogeneity and isotropy (so that only the radial modes matter). For example, the limits to the wavelengths of the vacuum modes can be given on the low end by the as-yet unknown characteristic length related to the quantum gravity (probably ~ the Planck length), and on the high end by the distance to the horizon. With reasonable estimates of these bounds, one gets values of the $\rho_{vac}$ comparable to the observed value [2,3]. Here we expand on and refine their model, and compare it with observations.

The GX formula can be written as:
$$\rho_{vac} = (h/16c)\,(L_{min}\,L_{max})^{-2}$$
$$= (\pi c^2/8G)\,(L_{Pl}/L_{min})^2\,L_{max}^{-2} \quad (1)$$
where $L_{min}$ and $L_{max}$ are the lower and the upper bounds to the vacuum modes, and $L_{Pl} = (Gh/2\pi c^3)^{-2}$

is the Planck length. (Throughout this paper we will be expressing $\rho_{vac}$ in mass density units). We will henceforth assume that $L_{min} = L_{Pl}$, or perhaps some multiple of $L_{Pl}$, and fixed in proper coordinates, but in the absence of a proper quantum theory of gravity, this reasonable assumption may not be exactly correct.

Note the *apparent* similarity to the Casimir energy density, $\rho_{cas} = h/ca^4$. However, there are essential differences: with a reasonable assumption that $L_{min} = L_{Pl}$, the Planck constant cancels from the new formula for $\rho_{vac}$, even though we are describing the contribution of quantum modes! Moreover, the GX energy density is absolute, and derived in a fundamentally different manner.

We can then express this in the form:
$$\Omega_{vac} = \rho_{vac}/\rho_{crit} = (\pi^2/3)(D_H/L_{max})^2 \quad (2)$$
$$\approx 3.29 (D_H/L_{max})^2$$
where $D_H = c/H$ is the Hubble radius. The same formula should apply at all redshifts. We note that unlike other forms of energy density which depend of the local densities of particles and their energies, this form of DE is qualitatively different: it depends on a global length scale, $L_{max}$.

The key question then is the nature of the upper bound $L_{max}$, and its change in redshift. We can come up with three viable choices, although it is possible that some other interpretation may be valid:

First, it may be fixed, $L_{max} = const. = D_{vac}$. This is essentially the cosmological constant model, where one mysterious number ($\Lambda$) is replaced by another ($D_{vac}$), whose physical meaning is yet to be determined (say, a coherence length of vacuum fluctuations?). For $\Omega_{vac} \approx 0.7$, $D_{vac} \approx 2.17 D_{H0}$, where $D_{H0}$ is the present value of the Hubble radius.

Second, it could be the distance to the horizon (i.e., to infinite redshift $z=\infty$), $D_\infty$, measured in proper units. It turns out that this does not work, as it leads to universes too young today, and EOS parameter $w \approx 0$, in a conflict with observations.

Finally, it could be the distance to the horizon measured in comoving units. This seems intuitively right, since the quantization should be done in commoving coordinates. Thus, we have:
$$L_{max} = D_\infty(0) = D_H \int_0^\infty dz/E(z) \quad (3)$$
where
$E(z) = [\Omega_m(1+z)^3 + \Omega_{rad}(1+z)^4 + \Omega_k(1+z)^2 + \Omega_{vac} f(z)]^{1/2}$,
and $f(z) = (1+z)^{3w+3}$ in the usual parametrization, is the redshift dependence of the DE. Henceforth we assume that we live in a spatially flat universe, $\Omega_k=0$, as strongly indicated by CMBR observations, and we assume the measured CMBR value $\Omega_{rad,0} = 1.08622 \times 10^{-5}$ in all computations.

In this model, the redshift dependence of the DE density, *as observed by us*, is:
$$\Omega_{vac}(z) = \Omega_{vac,0} f(z) = \Omega_{vac,0}[D_\infty(0)/D_\infty(z)]^2 \quad (4)$$
where $D_\infty(z)$ can be computed by replacing the lower limit of integration in eq.(3) with $z$. Therefore, we solve iteratively the integral equation which combines eqs. (2), (3), and (4), to obtain a unique, self-consistent solution for $\Omega_{vac}$ and its dependence with redshift. We get $\Omega_{vac,0} \approx 0.4883$, within 20% of the observed value, and with no free parameters whatsoever! This is a remarkable result, as it follows directly from the GX model, and the assumptions $L_{min}=L_{Pl}$, $L_{max}=D_\infty$ in commoving units, and $\Omega_k=0$.

While these assumptions are reasonable, they are not rigorously justified, and besides, the model in this simplest form does miss the observed value of $\Omega_{vac,0}$ by $\sim 20\%$. Thus, we introduce simple a parametrization of the form:
$$\Omega_{vac} = \kappa (\pi^2/3)(D_H/D_\infty)^2 \quad (5)$$
where the value of the phenomenological parameter $\kappa$ can be obtained from fits to observations. Its physical interpretation depends among other things on the exact understanding of quantum gravity, e.g., the minimum length may be some multiple of $L_{Pl}$, etc. It can be regarded as a "landscape parameter"; and so on. There is a unique correspondence between a value of $\kappa$, and the internally self-consistent value of $\Omega_{vac,0}$, as shown in Fig. 1.

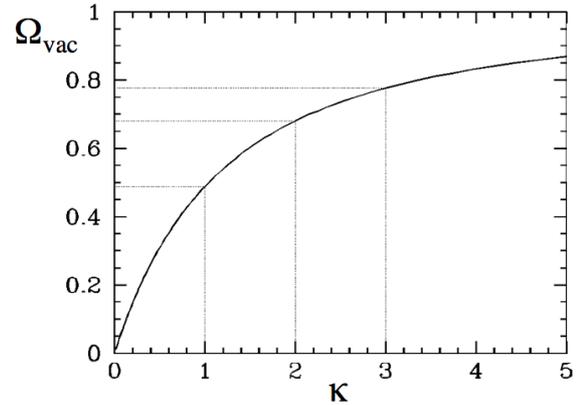

**Figure 1.** The correspondence between the value of the parameter $\kappa$, and the value of $\Omega_{vac,0}$.

It is apparent that values of $\kappa \sim 2 - 3$ will produce the observed value of $\Omega_{vac,0}$. Each value of $\Omega_{vac,0}$ in a self-consistent solution also produces dependence of distances and other quantities on redshift, enabling a comparison with the data.

## 2. Comparison with the observations

We use a combined set of distances to SN standard candles [6,7] as compiled by [8], to fit this model to the SN Hubble diagram (Fig. 2). The quality of the fit is indicated in Fig. 3, compared to the standard cosmological constant ($\Lambda$) model.

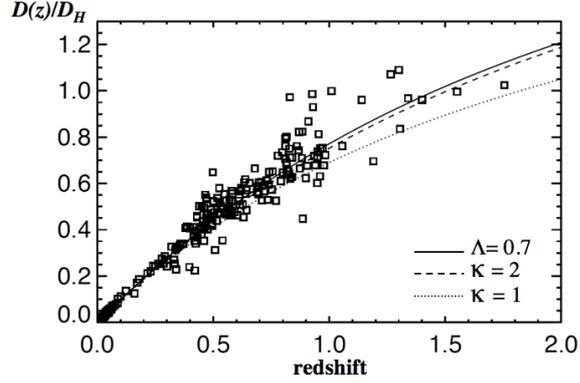

**Figure 2.** A Supernova Hubble diagram using linearized luminosity distances in Hubble units vs. redshift. The solid line shows the standard cosmological model with $\Omega_\Lambda = 0.7$, and dotted and dashed lines show the solutions for the vacuum fluctuations model for values of $\kappa = 1$ and 2.

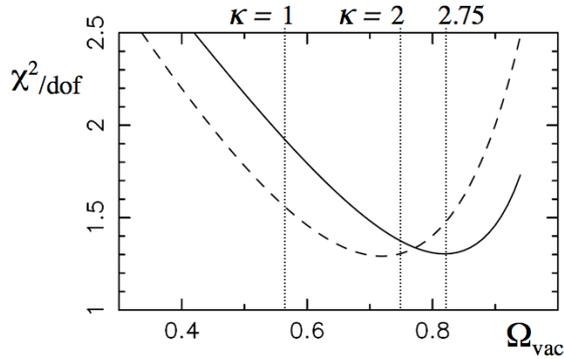

**Figure 3.** Reduced $\chi^2$ vs. $\Omega_{vac}$ for the fits to the Hubble diagram data shown in Fig. 2, for the standard $\Lambda$ model (dashed line), and the vacuum fluctuations model (solid line); the vertical dotted lines indicate the corresponding values of $\kappa$ from Fig. 1. Note that both models provide equally good optimal fit, as indicated by the depth of the $\chi^2$ curves.

Thus, the vacuum fluctuations model fits these data just as well as the cosmological constant model, but for a somewhat higher value of $\Omega_{vac,0}$, which is still fully compatible with all other currently available data.

Another important prediction is the value of the effective EOS parameter and its trend with redshift, $w(z)$, shown in Fig. 4. We compute this as the local power law slope of $f(z) = \Omega_{vac}(z)/\Omega_{vac,0} = (1+z)^{3w+3}$ in our self-consistent solutions.

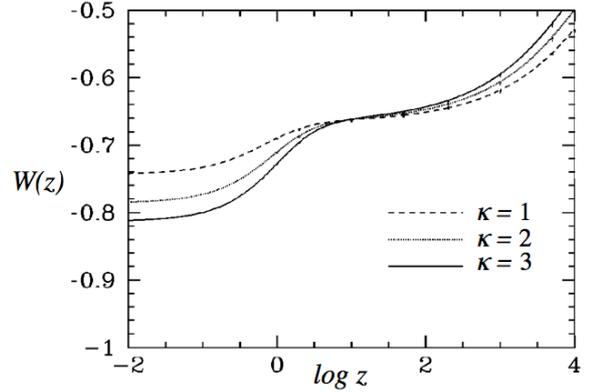

**Figure 4.** Values of the effective EOS parameter $w(z)$ vs. redshift for three values of $\kappa$ as indicated with different line types.

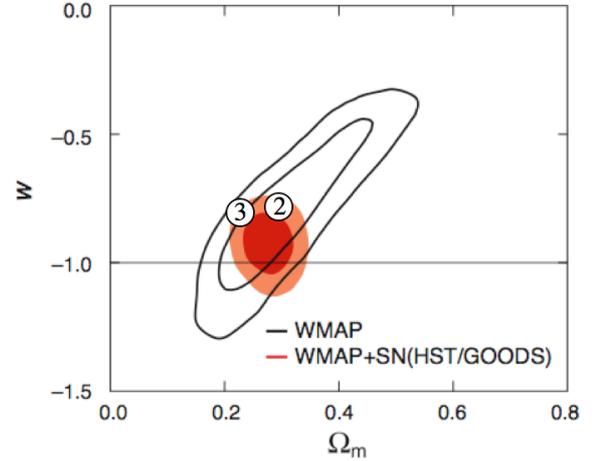

**Figure 5.** Constraints on the EOS parameter $w$ and $\Omega_m$, from [9]. Values for our $z = 0$ solutions for the values of $\kappa = 2$ and 3 are shown in small circles.

We get the present values of $w \approx -0.8$, which is compatible with the current determinations, as shown, e.g., in Fig. 5. We note that this indicative comparison is not entirely correct, as the probability contours shown in Fig. 5 (and all other similar published plots) are computed assuming $w = const.$, whereas in our model $w$ changes with redshift.

Finally, we explore the question on whether this model is in conflict with the well established CMBR and cosmic nucleosynthesis results, given that $\Omega_{vac}$ increases with redshift. Fig. 6 shows the relative values of different density components out to $z \sim 10^9$. We see that the DE contribution is still negligible at the recombination and the nucleosynthesis eras, and thus there is no conflict with the early universe which is probed by observations.

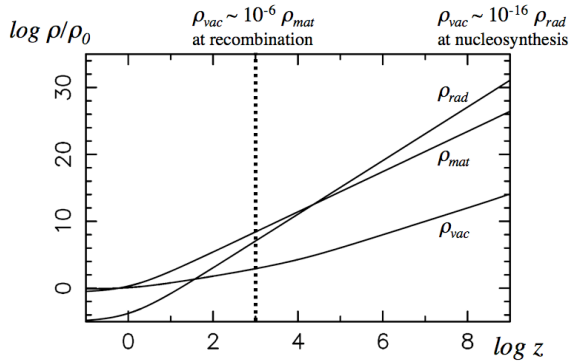

**Figure 6.** Behavior of the various energy density components as functions of redshift, normalized by the present day critical density.

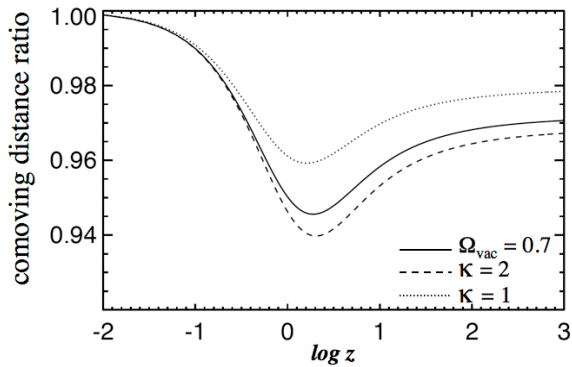

**Figure 7**. Ratio of the comoving distance to a given redshift computed in the vacuum fluctuations model, and the distance computed in the standard $\Lambda$ model for the same value of $\Omega_{vac}$ today.

Thus, the model with the parameter $\kappa$ in the range $\sim 2 - 3$ fits all of the currently available data as well as the cosmological constant model for the observed values of $\Omega_{vac} \sim 0.7 - 0.8$. How can we test it further?

Fig. 7 shows the ratio of the comoving distances computed for this model and the standard $\Lambda$ model, for the same values of $\Omega_{vac,0}$. This maps into different model predictions for various cosmological tests. As we can see, the differences are modest, at most at a $\sim 5\%$ level, but can be presumably tested by the future precision cosmology experiments. The differences are maximized, with a characteristic dip shape, in the redshift range $z \sim 0.5 - 5$, which is about right for the forthcoming data sets on SNe and SZ measurements of clusters, etc.

## 3. Concluding comments

We note that the model (assuming that the minimum scale is indeed fixed at $L_{min} \sim L_{Pl}$) really predicts that $\rho_{vac} G/c^2 \sim L_{max}^{-2}$, and thus in principle some change in redshift could be attributed to a variation in the fundamental constants $c$ and $G$ – for which there are excellent limits from various measurements. We find this possibility too speculative for now, but such models have been explored by [10].

Several other discussions of models based in the GX theory include [11,12,13,14].

It is also interesting to note that essentially the same expression for the DE density, and the same uncertainty as to the choice of the maximum length $L_{max}$ is found in various holographic models (see, e.g., [5,15,16] and references therein), even though the physical motivation and approach there are completely different from the vacuum fluctuations model. While we cannot enter here into a discussion of these models, this apparent coincidence is intriguing, and reminds us of the Heisenberg and Schrödinger approaches to quantum mechanics, both of which described equally well the same underlying physical reality. Whether the same turns out to be the case here, only the future will show.

If the observed value of the DE density can be accounted for just by the (inevitable) quantum fluctuations of the ground state of the physical vacuum, the implication is that the value of the ground state is indeed zero – or certainly consistent with zero. This effectively removes the fine-tuning problem which plagues many proposed models for DE. Of course, we cannot tell why there is such a perfect cancellation, and this remains as a challenge for the theory.

The vacuum fluctuations model is based on some well established physics, plus a few reasonable assumptions. It is in an excellent agreement with all of the available data, and it makes testable predictions. These are very positive features. However, there is a fundamental problem which we must address next.

In our computations, we evaluated the DE density as a function of redshift by using the value of the commoving distance to the horizon *there, as seen by us*, which is the correct procedure for evaluating observables in our frame of reference. However, since this (vacuum fluctuationa) energy density depends on a square of a length, which is not invariant, a hypothetical observer at that redshift will derive a different value of $\rho_{vac}$ there. In fact, every observer at all redshifts will derive the same value of $\Omega_{vac}$ in their frame of reference, as they would do exactly the same computation as we did.

How to resolve this apparent paradox? One possibility is that our choice of $L_{max}$ as the commoving distance to the horizon is incorrect, even though it produces such excellent fits to the observations. A simple solution would be to declare $L_{max} = D_{vac} = const.$, which is the only invariant solution, i.e., to revert to the cosmological constant model, thus leaving us with a mystery of the physical meaning of $D_{vac}$. Another possibility is to allow for a change in redshift of the lower bound, $L_{min}$, but that clearly reaches into an unknown domain of the physics on the Planck scale and is thus purely *ad hoc* at this time. And finally, it is possible that there is something fundamentally relativistic (or even new) that we are missing in our approach when looking at the DE in different frames of reference. (For a related discussion of invariants in the context of GX model see, e.g., [14]). Further discussion of these issues is beyond the scope of the present contribution, and will be presented in a forthcoming paper.

SGD acknowledges a partial support from the NSF grant AST-0407448 and the Ajax Foundation. We thank numerous colleagues for useful discussions.